\documentclass[sigconf]{acmart}
\AtBeginDocument{%
  }
\usepackage{graphicx}
\usepackage{tikz}
\usetikzlibrary{positioning}
\setcopyright{none}
\begin{document}

\title{Time Semantics and Liveness Artifacts in Adversarial Consensus Simulation}


\author{Nasit S Sony}
\affiliation{%
  \institution{Independent Researcher}
  \city{Merced}
  \state{CA}
  \country{USA}
}
\email{nasitsony96@gmail.com}
\author{Xianzhong Ding}
\affiliation{%
  \institution{University of California, Merced}
  \city{Merced}
  \state{CA}
  \country{USA}
}
\email{xding5@ucmerced.edu}

\renewcommand{\shortauthors}{Sony}

\begin{abstract}
Adversarial consensus simulators require not only a model of message delivery but also a model of time. For timeout-sensitive protocols, coupling protocol-time progression to scheduler activity can allow communication scheduling to influence when failure detectors expire.

We study this interaction in VeriProtocol using controlled Multi-Paxos and Raft experiments. In the primary Multi-Paxos study, we compare EventCoupled timing, where every scheduler opportunity advances protocol time, with RoundTick timing, which decouples individual scheduler events from timeout progression. At N=5, increasing the bounded message-delay budget from K=6 to K=14 increased the observed EventCoupled leadership-instability rate from 5\% to 100\% across the tested grid, while no additional valid leader elections occurred in any of the corresponding 140 RoundTick executions. At K=0, neither model produced additional elections.

Matched-trace analysis identifies the mechanism: under EventCoupled timing, scheduler activity used to perturb communication also directly advances failure-detector state. Under RoundTick, comparable scheduler activity does not imply equivalent protocol-time progression. A targeted heartbeat-delay control confirms that RoundTick does not suppress timeout-driven elections; an additional election remains reachable when leader contact is unavailable across sufficient logical time.

A deterministic Raft experiment provides cross-protocol validation. Every tested nonzero heartbeat-delay budget produced additional successful elections under EventCoupled, whereas none did under RoundTick. This difference persisted when RoundTick was extended to 400 logical ticks, matching the EventCoupled logical-time observation horizon.

These results identify a methodological hazard in adversarial consensus simulation: coupling communication scheduling to protocol-time progression can implicitly expand the capability of an adversary intended to control communication alone. Evaluations of timeout-sensitive distributed protocols should therefore report their time semantics explicitly and treat communication scheduling and protocol-time advancement as distinct experimental dimensions.

\end{abstract}



\maketitle

\section{Problem and Motivation}

Adversarial schedulers are useful for evaluating the liveness behavior of distributed consensus protocols by systematically delaying, reordering, or prioritizing protocol messages. Such simulations, however, require not only a model of message delivery but also a model of time. This distinction becomes important for timeout-driven protocols: if protocol time advances with scheduler activity, the number of simulated message-delivery events can inadvertently determine when failure detectors expire.

We encountered this issue while evaluating Multi-Paxos under bounded adversarial message scheduling. Under an event-coupled time model, targeted delay of AcceptRequest traffic appeared to induce substantial post-recovery leadership instability. The effect persisted after replacing a global network queue with per-sender scheduling, suggesting that global serialization was not its primary cause.

However, when timeout progression was decoupled from individual scheduler events, the apparent instability disappeared across the matched Multi-Paxos configurations. A subsequent Raft experiment exhibited the same qualitative dependence on time semantics. These observations suggest a broader methodological concern: coupling scheduler activity to protocol-time progression can implicitly expand the capability of an adversary that was intended to control communication alone. This led us to ask:

\section{ Research Question}

How can the coupling between scheduler activity and protocol-time progression affect liveness conclusions in adversarial consensus simulation?

\section{ Simulator and Time Models}

VeriProtocol is a deterministic consensus simulation framework in which protocol messages are placed into a simulated network and delivered according to a configurable scheduler. The scheduler can perturb execution by delaying or selecting messages, while seeded executions provide reproducible schedules for controlled comparisons.

The primary study considers stable Multi-Paxos with leader failure and recovery. Followers maintain heartbeat-based failure detectors and initiate a new election when their heartbeat age reaches an election-timeout threshold. We additionally use a timeout-sensitive Raft scenario as cross-protocol validation of the observed mechanism. Across these experiments, we compare two semantics for advancing simulated protocol time.

\subsection{EventCoupled}

Under EventCoupled semantics, each scheduler step advances protocol time. Consequently, message deliveries, adversarial delay decisions, and other scheduler activity also advance heartbeat and timeout state. Scheduler activity can therefore accelerate failure-detector progression even when the number of scheduler events does not correspond to an equivalent amount of elapsed time.

\subsection{RoundTick}

Under RoundTick semantics, protocol-time progression is decoupled from individual scheduler events. For a configuration with $N$ nodes, one logical tick occurs after $N$ scheduler opportunities, and heartbeat and failure-detector timers advance only on these logical ticks. RoundTick therefore removes the direct proportionality between scheduler-event count and protocol-time progression.

This distinction matters because consensus liveness and failure
detection depend on timing assumptions about communication and
execution~\cite{dwork1988partial,chandra1996failure}. Wall-clock
failure detectors expire according to elapsed time rather than
according to the number of messages delivered or scheduling decisions
performed. RoundTick is therefore more faithful along the specific dimension studied here: scheduler activity alone does not cause protocol time to advance at the same rate as under EventCoupled semantics. We do not claim that RoundTick exactly reproduces physical wall-clock execution; it remains a deterministic logical-time abstraction that corrects this particular event-time coupling.

The two models enable controlled comparisons in which the protocol, scheduler, network model, and workload are held fixed while changing how scheduler progress is translated into protocol time. For the seeded Multi-Paxos comparisons, the random seed is also held fixed.

RoundTick does not disable timeout-driven elections. As demonstrated by the heartbeat-delay control in Section~6, an additional election remains reachable under RoundTick when leader contact is unavailable across sufficient logical-time progression.

\section{  Experimental Setup}

We evaluate a Multi-Paxos recovery scenario in which the initial leader fails and the remaining replicas must elect a replacement while preserving previously accepted values. The adversarial scheduler targets Multi-Paxos AcceptRequest traffic using a bounded delay budget $K$. Safety is checked throughout each execution, and recovery is considered stable when a heartbeat from the current highest-ballot post-failure leader is successfully delivered.

For Multi-Paxos, we define leadership instability as at least one additional valid leader election or ballot advance after the expected post-failure recovery election. For Raft, the first successful election establishes the initial leader, so we define instability as more than one successful leader election. The baselines differ because the experimental scenarios differ: Multi-Paxos begins with a leader-failure-and-recovery event, whereas the Raft experiment begins with initial leader establishment. In both cases, the measured event is an additional successful leadership change beyond the election expected by the scenario. Transient timeout observations that do not produce an additional valid election are not classified as instability.

We initially evaluated a global network queue, then introduced a per-sender network model to test whether observed instability resulted from serialization of otherwise independent communication paths. At $N=5$, $K=11$, instability decreased only modestly from 85\% under the global queue to 75\% under per-sender scheduling, indicating that global serialization alone did not explain the effect.

We subsequently compared EventCoupled and RoundTick using the per-sender network model. For the Multi-Paxos seed-based experiments, each evaluated configuration uses 20 deterministic seeds unless otherwise noted.

\section{Results}

\subsection{Time semantics qualitatively change the observed stability result}

We first compare the two time models at $N=5$, $K=11$, using the per-sender network model and the same 20 deterministic seeds. Under EventCoupled timing, 15 of 20 executions (75\%) exhibited leadership instability after the expected recovery election. The average number of view changes was 2.95, with a maximum of four.

Under RoundTick timing, none of the 20 executions exhibited an additional valid leader election. The instability rate therefore changed from 75\% to 0\%, while the average number of view changes decreased from 2.95 to 1.00. All executions preserved the expected Multi-Paxos safety property.

\begin{table}[t]
\centering
\caption{Comparison of EventCoupled and RoundTick time semantics for $N=5$, $K=11$ over 20 seeds.}
\label{tab:time-model-comparison}
\resizebox{\columnwidth}{!}{%
\begin{tabular}{lrrrrrrrr}
\toprule
Time Model & Runs & Unstable & Instability & Avg. Views & Max Views & Max Ballot & Stable Recovery & Avg. Stable Tick \\
\midrule
EventCoupled & 20 & 15 & 75\% & 2.95 & 4 & 5 & 20/20 & 231.05 \\
RoundTick    & 20 & 0  & 0\%  & 1.00 & 1 & 2 & 20/20 & 37.35 \\
\bottomrule
\end{tabular}%
}
\end{table}

Because the protocol, scheduler, network model, workload, and seeds are held fixed, this comparison changes only the simulator's protocol-time semantics. The previously observed leadership churn therefore does not persist when protocol-time progression is decoupled from individual scheduler events.

\subsection{The time-semantics effect persists across delay budgets}

We next varied the adversarial delay budget at $N=5$ using the per-sender network model, evaluating
$K \in \{0,6,8,10,11,12,14\}$
across the same 20 deterministic seeds under each time model. At $K=0$, neither model produced additional leader elections, providing a no-delay control. Across the tested grid, the EventCoupled instability rate increased with $K$: 5\% of executions were unstable at $K=6$, 40\% at $K=8$, 50\% at $K=10$, 75\% at $K=11$, 85\% at $K=12$, and 100\% at $K=14$.

Under RoundTick, no additional valid leader elections occurred at any tested delay budget. All 140 RoundTick executions remained stable, while average stable-recovery time increased modestly from 35.00 logical ticks at $K=0$ to 38.00 at $K=14$. Thus, decoupling individual scheduler events from timeout progression did not eliminate the effect of adversarial scheduling on recovery latency; rather, it changed whether the tested perturbations also produced additional leader elections.

We additionally evaluated matched configurations at $N=3$ and $N=7$
to determine whether the time-semantics effect was specific to the
$N=5$ configuration. At $N=3$, $K=16$, 14 of 20 EventCoupled
executions exhibited additional leader elections, whereas none of the
20 matched RoundTick executions did. At $N=7$, $K=10$, all 20
EventCoupled executions exhibited additional leader elections, compared
with zero of 20 under RoundTick. Stable recovery was observed within
the 400-scheduler-step window in all 20 RoundTick executions at both
node counts; under EventCoupled it was observed in all 20 executions
at $N=3$ but in none of the 20 executions at $N=7$.

\begin{table*}[t]
\centering
\caption{Cross-node-count validation under matched EventCoupled and
RoundTick configurations. Stable recovery reports executions in which
stable recovery was observed within the 400-scheduler-step observation
window; the average recovery tick is computed only over executions
where stable recovery was observed.}
\label{tab:cross-n}
\small
\begin{tabular}{rrlrrrrrr}
\toprule
$N$ & $K$ & Model & Unstable & Avg. VC & Max VC & Max ballot &
Stable & Avg. recovery tick \\
\midrule
3 & 16 & EventCoupled & 14/20 (70\%)  & 1.70 & 2 & 3 & 20/20 & 110.85 \\
3 & 16 & RoundTick    & 0/20 (0\%)    & 1.00 & 1 & 2 & 20/20 & 41.40  \\
5 & 10 & EventCoupled & 10/20 (50\%)  & 2.30 & 5 & 6 & 20/20 & 193.60 \\
5 & 10 & RoundTick    & 0/20 (0\%)    & 1.00 & 1 & 2 & 20/20 & 37.05  \\
7 & 10 & EventCoupled & 20/20 (100\%) & 6.00 & 6 & 7 & 0/20  & --     \\
7 & 10 & RoundTick    & 0/20 (0\%)    & 1.00 & 1 & 2 & 20/20 & 34.00  \\
\bottomrule
\end{tabular}
\end{table*}

These cross-$N$ configurations are not an $N$-scaling experiment because the evaluated delay budgets differ across cluster sizes. They provide a narrower validation: the observed dependence on time semantics is not confined to the $N=5$ configuration.

The RoundTick result does not imply that adversarial delay has no effect.
In the $N=5$ sweep, recovery latency still changes with $K$. Rather, the
qualitative effect changes: bounded message delay can slow recovery without
necessarily inducing additional valid leader elections. This distinction
suggests that the earlier leadership-instability result was produced by the
interaction between adversarial scheduling and event-coupled timeout
progression, rather than by the targeted message delay alone.

\section{Matched-Trace Mechanism}

To understand why the two time models produce different leadership outcomes, we examine a matched execution using seed 42 at $N=5$ and $K=11$. The protocol configuration, workload, adversarial scheduler, per-sender network model, and random seed are identical across the two executions; only the semantics of protocol-time progression differ. This comparison allows us to trace how scheduler activity interacts with heartbeat generation and follower timeout state under EventCoupled and RoundTick.

The matched execution highlights that the difference is not simply a reduction in scheduler activity. In the seed-42 comparison, EventCoupled reached stable recovery at scheduler step 176, after 175 logical ticks and after an additional election raised the ballot to 3. RoundTick, by contrast, required 191 scheduler steps to reach stable recovery but advanced through only 38 logical ticks and remained at ballot 2. Thus, RoundTick performed more raw scheduler work before stable recovery while avoiding the additional election. This contrast provides direct evidence for the mechanism: under EventCoupled, scheduler activity itself contributes directly to timeout progression, whereas under RoundTick the same activity does not imply an equivalent amount of elapsed protocol time.

The heartbeat-delay control distinguishes time decoupling from suppression of timeout-driven elections. Under RoundTick, we used a bounded scheduler that repeatedly postponed ballot-2 heartbeats from the live recovery leader (node 2) to a specific follower (node 3). With a follower timeout threshold of 20 logical ticks and a heartbeat-delay budget of 150 scheduler opportunities, the execution produced an additional valid election and advanced from ballot 2 to ballot 3. Thus, RoundTick does not prevent adversarially induced elections. Rather, additional elections require leader contact to remain unavailable across sufficient logical-time progression instead of allowing scheduler activity itself to advance the follower toward its timeout at the EventCoupled rate.

\begin{figure*}[t]
\centering
\begin{tikzpicture}[
    node distance=0.45cm,
    box/.style={
        draw,
        rounded corners,
        align=center,
        text width=3.7cm,
        minimum height=0.75cm,
        font=\small
    },
    result/.style={
        draw,
        rounded corners,
        align=center,
        text width=3.7cm,
        minimum height=0.75cm,
        font=\small\bfseries
    },
    arrow/.style={->, thick}
]

\node[box] (ec1) at (0,0)
    {\textbf{EventCoupled}\\scheduler activity};
\node[box, below=of ec1] (ec2)
    {Scheduler events advance\\protocol time};
\node[box, below=of ec2] (ec3)
    {Failure-detector state\\reaches timeout};
\node[result, below=of ec3] (ec4)
    {Additional election};

\draw[arrow] (ec1) -- (ec2);
\draw[arrow] (ec2) -- (ec3);
\draw[arrow] (ec3) -- (ec4);

\node[box] (rt1) at (5.0,0)
    {\textbf{RoundTick}\\scheduler activity};
\node[box, below=of rt1] (rt2)
    {Scheduler events do not advance\\protocol time one-for-one};
\node[box, below=of rt2] (rt3)
    {Insufficient logical-time\\progression};
\node[result, below=of rt3] (rt4)
    {No additional election};

\draw[arrow] (rt1) -- (rt2);
\draw[arrow] (rt2) -- (rt3);
\draw[arrow] (rt3) -- (rt4);

\node[box] (ctl1) at (10.0,0)
    {\textbf{RoundTick}\\targeted heartbeat postponement};
\node[box, below=of ctl1] (ctl2)
    {Leader contact unavailable across\\sufficient logical time};
\node[box, below=of ctl2] (ctl3)
    {Timeout threshold\\reached};
\node[result, below=of ctl3] (ctl4)
    {Additional election};

\draw[arrow] (ctl1) -- (ctl2);
\draw[arrow] (ctl2) -- (ctl3);
\draw[arrow] (ctl3) -- (ctl4);

\end{tikzpicture}

\caption{Mechanism and control for the time-semantics effect.
EventCoupled allows scheduler activity to contribute directly to timeout
progression. RoundTick removes this one-for-one coupling, but timeout-driven
elections remain reachable when leader contact is unavailable across
sufficient logical time.}
\label{fig:time-semantics-mechanism}
\end{figure*}
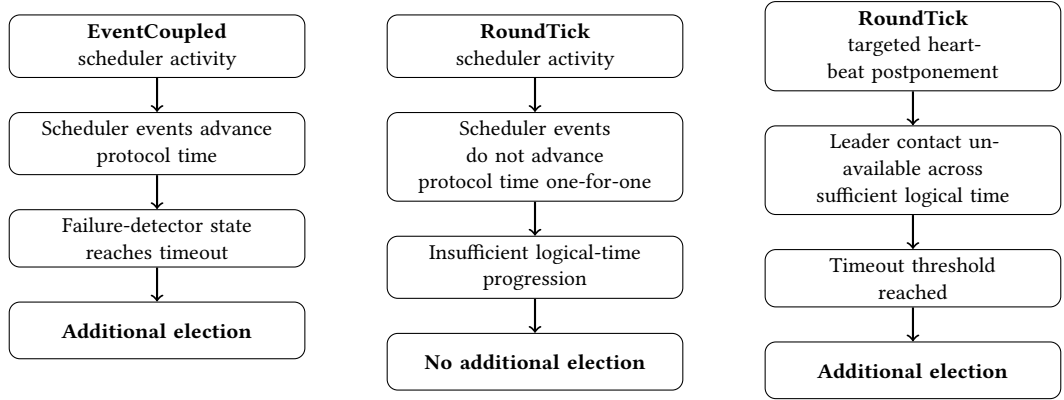

\section{Cross-Protocol Validation with Raft}

To test whether the observed time-semantics effect was specific to Multi-Paxos, we constructed a timeout-sensitive Raft experiment using five nodes, the per-sender network model, and a bounded scheduler that targets term-1 heartbeats from the initial leader to its followers. Heartbeats are generated every five logical ticks, and follower election timeouts are deterministically staggered at 20, 22, 24, 26, and 28 logical ticks. We evaluated delay budgets
$K \in \{0,5,10,15,20,25,30\}$.
Because the Raft scheduler and network configuration contain no randomized choices, repeated executions of a fixed configuration produce the same trace. We therefore record one execution per configuration; unlike the seeded Multi-Paxos experiments, there is no random variable over which repeated runs would provide additional sampling evidence.

We classify a Raft execution as unstable when more than one successful leader election occurs: the first election establishes the initial leader, while any subsequent successful election represents additional leadership change. At $K=0$, both EventCoupled and RoundTick remained stable, with one successful election and no follower timeouts. Under EventCoupled, every tested nonzero delay budget,
$K \in \{5,10,15,20,25,30\}$,
produced at least one additional successful leader election. Election counts ranged from two to three across these configurations. Under RoundTick, all tested delay budgets remained at one successful election with no follower timeouts.

We additionally evaluated two observation horizons to test whether the RoundTick result could be explained simply by a shorter logical-time observation window. Under the scheduler-step horizon, both models received 400 scheduler opportunities, corresponding to 400 logical ticks under EventCoupled but only 80 logical ticks under RoundTick. We therefore repeated the RoundTick experiment for 400 logical ticks, requiring 2000 scheduler opportunities. Despite receiving five times as many scheduler opportunities and the same 400-logical-tick observation horizon as EventCoupled, RoundTick still produced no additional successful elections at any tested delay budget. This control rules out the shorter logical-time observation horizon as the explanation for the absence of additional elections under RoundTick in this experiment.

\begin{table}[t]
\centering
\caption{Raft validation under scheduler-step and logical-time observation horizons.}
\label{tab:raft-validation}
\resizebox{\columnwidth}{!}{%
\begin{tabular}{llrrll}
\toprule
Time Model & Observation Horizon & Scheduler Steps & Logical Ticks & Tested $K$ & Additional Elections \\
\midrule
EventCoupled & Scheduler steps & 400  & 400 & 0     & No \\
EventCoupled & Scheduler steps & 400  & 400 & 5--30 & Yes at every tested $K$ \\
RoundTick    & Scheduler steps & 400  & 80  & 0--30  & No \\
RoundTick    & Logical time    & 2000 & 400 & 0--30  & No \\
\bottomrule
\end{tabular}%
}
\end{table}

The Raft experiment provides cross-protocol evidence for the same simulator-level mechanism observed with Multi-Paxos. Under EventCoupled, scheduler activity used to postpone communication also directly advances timeout-related protocol time. Under RoundTick, individual scheduler opportunities no longer have this direct effect. The Raft and Multi-Paxos experiments use different protocol scenarios and should not be interpreted as a quantitative comparison of their susceptibility to adversarial scheduling. Rather, they demonstrate that the time-semantics effect is not confined to the Multi-Paxos recovery scenario.

\section{Methodological Implications}

Our results show that time semantics are part of the experimental model in adversarial consensus simulation, rather than an implementation detail. In particular, coupling timeout progression to individual scheduler events changes the effective capability of the modeled adversary. A scheduler specified to control message ordering and delay can also indirectly control how quickly protocol failure detectors perceive time to be passing, even though control over protocol time is not explicitly part of the adversary definition.

The Multi-Paxos experiments demonstrate the behavioral consequence of this additional capability. Under EventCoupled semantics, bounded perturbations of protocol traffic produced additional leader elections at multiple tested nonzero delay budgets, including 75\% of executions at $N=5$ and $K=11$. Under matched RoundTick configurations, these additional elections disappeared. The matched seed-42 trace further showed that this difference was not caused by RoundTick performing less scheduler work: RoundTick required more scheduler steps before stable recovery while advancing through substantially fewer logical ticks and avoiding the additional election. The Raft validation exhibited the same qualitative dependence on time semantics: every tested nonzero heartbeat-delay budget produced additional successful elections under EventCoupled, whereas none did under RoundTick, including when both models were observed for 400 logical ticks.

These observations suggest a methodological requirement for adversarial liveness evaluation: the capabilities assigned to the communication adversary should be separated from the mechanism by which protocol time advances. Otherwise, an adversary intended to manipulate communication can acquire an implicit timing capability through the simulator's execution semantics. Apparent liveness degradation may then reflect the combination of the intended communication perturbation and this additional timing capability, rather than the communication perturbation alone.

More generally, evaluations of timeout-sensitive distributed protocols should report their time semantics explicitly. Results involving leader elections, retries, failure detectors, leases, or other timer-driven behavior may depend on whether simulated time advances with events, rounds, or an independent clock. Reproducible adversarial schedules are therefore not sufficient by themselves; the mapping from scheduler execution to protocol time is also part of the experimental specification.
\section{Limitations}

This study evaluates time semantics within deterministic simulations of Multi-Paxos and Raft rather than deployments using physical clocks and real network delays. RoundTick should therefore not be interpreted as an exact representation of wall-clock execution. It is a controlled abstraction that removes the direct coupling between individual scheduler events and protocol-time progression. Our conclusion is consequently about the sensitivity of simulated liveness behavior to this coupling, rather than a claim that RoundTick reproduces all timing behavior of deployed systems.

The experimental scope covers stable Multi-Paxos and a timeout-sensitive Raft scenario, both within the same deterministic simulator. The Raft experiment provides cross-protocol evidence that the observed time-semantics effect is not confined to the Multi-Paxos recovery scenario, but it does not establish that the effect occurs in all timeout-sensitive consensus protocols or under all workloads and adversarial strategies. Generalization to protocols such as PBFT, HotStuff, and other timer-driven distributed systems remains future work.

The experiments use specific bounded adversarial schedulers: targeted AcceptRequest delay in the primary Multi-Paxos study and targeted heartbeat delay in the Raft validation. Other adversaries, including partitions, correlated delays, adaptive scheduling strategies, or perturbations of different protocol message classes, may produce different behavior under either time model. Our results should therefore not be interpreted as showing that RoundTick prevents adversarially induced leadership changes; the heartbeat-delay control in Section 6 explicitly demonstrates that additional elections remain reachable under RoundTick when leader contact is unavailable across sufficient logical time.

Finally, RoundTick advances logical time at a fixed relationship to scheduler opportunities. This removes the one-event-one-time-unit coupling of EventCoupled, but it remains a discrete simulation policy rather than an independently modeled physical clock. Future work could evaluate alternative decoupled clock models, including explicit timer events or simulated wall-clock advancement, to determine how sensitive the observed results are to the particular decoupling strategy.

\section{Threats to Validity}

One threat concerns the implementation of the two time models. EventCoupled and RoundTick are implemented within the same simulator, so implementation errors could affect the comparison. We mitigate this risk through controlled comparisons that hold the protocol, scheduler, network model, and workload fixed while changing the rule governing timeout progression; for the seeded Multi-Paxos comparisons, we additionally hold the random seed fixed.

A second threat concerns the definition of leadership instability. For Multi-Paxos, we classify an execution as unstable when the expected recovery election is followed by at least one additional valid leader election and ballot advance. For Raft, we classify an execution as unstable when the initial leader-establishing election is followed by at least one additional election. In both cases, the metric therefore captures additional election activity beyond the protocol-specific baseline rather than counting the baseline election itself. For Multi-Paxos, timeout observations from replicas that retain stale leader state are not counted as instability unless they result in a valid additional election.

A third threat is that RoundTick might suppress timeout behavior generally rather than specifically remove event-time coupling. The heartbeat-delay control in Section 6 addresses this alternative explanation by demonstrating that timeout-driven additional elections remain reachable under RoundTick when leader contact remains unavailable across sufficient logical-time progression.

Finally, the evaluated executions cover a finite set of cluster sizes, delay budgets, and schedules. The Multi-Paxos experiments use multiple seeded executions for each evaluated configuration, while the Raft validation uses deterministic executions under the specified heartbeat-delay schedule. These experiments establish the phenomenon within the evaluated configuration space but do not exhaust all possible schedules or adversarial strategies. We therefore treat the results as evidence of a specific simulator-modeling hazard rather than a universal characterization of consensus liveness.

\section{Related-Work Positioning}

Prior work provides several complementary approaches to testing and
simulating distributed systems. Randomized and adversarial testing
techniques explore unfavorable schedules and injected faults, including
randomized schedule exploration~\cite{ozkan2018randomized} and Byzantine
fault injection~\cite{winter2023byzzfuzz}. Deterministic and failure-oriented testing similarly enables
reproducible evaluation under controlled failures, message delays,
timeouts, and varied execution schedules
~\cite{chandra2007paxos,zhou2021foundationdb}. Recent frameworks such as
QUANTAS 2~\cite{oglio2026quantas2} and
DScale~\cite{shprenger2026dscale} illustrate distinct simulation
abstractions for evaluating distributed algorithms and systems.

Time representation has also been studied explicitly in distributed
testing. Blom et al.\ develop semantics for simulated time in
host-based testing and examine timed behavior under simulated rather
than physical time~\cite{blom2008simulated}. This prior work reinforces
that simulated-time semantics are part of the testing model; our study
examines how such semantics interact specifically with adversarial
scheduling and timeout-sensitive consensus liveness.

Our result should not be interpreted as showing that existing
distributed-systems simulators generally couple scheduler events to
protocol time. Simulation frameworks can adopt different execution and
timing abstractions; for example, QUANTAS uses an explicit round-based
execution model~\cite{oglio2026quantas2}. Instead, our concern is
conditional: when a simulator advances timeout-related protocol time as
a direct consequence of scheduler-event progression, that choice can
interact with an adversarial scheduler in a way that changes the
effective capabilities of the modeled adversary.

This work therefore focuses on a complementary methodological question:
whether communication scheduling and protocol-time progression are
independently controlled experimental dimensions. Through matched
EventCoupled and RoundTick executions, we show that coupling these
dimensions can qualitatively change observed leadership behavior while
the protocol, communication adversary, network model, workload, and seed
remain fixed. The Raft validation provides cross-protocol evidence for
the same simulator-level mechanism.

The contribution is consequently not a new consensus protocol,
scheduler, failure detector, or general-purpose simulation framework.
It is a controlled empirical demonstration that simulator time semantics
can interact with adversarial schedule exploration to alter
timeout-sensitive liveness conclusions, together with the methodological
implication that the communication adversary's explicit capabilities
should be distinguished from implicit influence over protocol-time
progression.
\paragraph{Artifact availability.}
The simulator, experiment configurations, and result artifacts used in
this study are publicly available at
\url{https://github.com/NasitSony/veriprotocol}.
\section{ Conclusion}

This study began with an apparent Multi-Paxos liveness result: bounded adversarial delay of AcceptRequest traffic produced repeated post-failure leader elections. Initial validation showed that the effect persisted after replacing a globally serialized network queue with per-sender scheduling, suggesting that queue serialization alone did not explain the instability. A controlled comparison of time semantics revealed a more fundamental mechanism: the observed leadership behavior depended strongly on how scheduler execution was mapped to protocol-time progression.

Under EventCoupled semantics, where each scheduler opportunity advances protocol time, bounded communication perturbations produced additional leader elections at multiple tested nonzero-delay Multi-Paxos configurations. At $N=5$ and $K=11$, $15$ of $20$ executions exhibited additional elections. Under matched RoundTick configurations, no additional elections occurred. Matched-trace analysis showed that RoundTick could perform more scheduler work before stable recovery while accumulating substantially less logical time, and a targeted heartbeat-delay control confirmed that timeout-driven additional elections remain reachable under RoundTick when leader contact is unavailable across sufficient logical time.

The Raft validation provided cross-protocol evidence for the same simulator-level mechanism. Every tested nonzero heartbeat-delay budget produced additional successful elections under EventCoupled, whereas none did under RoundTick. This result persisted when both models were observed for the same 400 logical ticks, ruling out the shorter logical-time horizon as an explanation for the RoundTick result in this experiment.

The resulting lesson is methodological rather than a claim about the vulnerability of either Multi-Paxos or Raft. Coupling scheduler-event progression to protocol-time progression changes the effective capability of a communication adversary: message-scheduling activity can also indirectly accelerate timeout progression. Adversarial evaluations of timeout-sensitive distributed protocols should therefore treat communication scheduling and protocol-time advancement as distinct experimental dimensions, report their time semantics explicitly, and avoid attributing liveness degradation solely to communication perturbations when the simulator also grants the scheduler implicit influence over protocol time.
\bibliographystyle{ACM-Reference-Format}
\bibliography{sample-base}

\appendix
\end{document}